\documentstyle[aps,preprint,floats,epsfig,epsf]{revtex}
\begin{document}
\tighten

\title{Domain-Wall Induced Quark Masses \\
  in Topologically-Nontrivial Background}
\author{Valeriya Gadiyak, Xiangdong Ji, Chulwoo Jung}
\bigskip

\address{
Department of Physics \\
University of Maryland \\
College Park, Maryland 20742 \\
{~}}

\date{UMD PP\#00-055 ~~~DOE/ER/40762-201~~~ February 2000}

\maketitle

\begin{abstract}
In the domain-wall formulation of chiral fermion, the 
finite separation between domain-walls ($L_s$) induces 
an effective quark mass ($m_{\rm eff}$) which complicates 
the chiral limit. In this work, we study the size of the 
effective mass as the function of $L_s$ and the domain-wall 
height $m_0$ by calculating the smallest eigenvalue
of the hermitian domain-wall Dirac operator in 
the topologically-nontrivial background fields. 
We find that, just like in the free
case, $m_{\rm eff}$ decreases exponentially in $L_s$ 
with a rate depending on $m_0$. However, quantum 
fluctuations amplify the wall effects significantly.
Our numerical result is consistent with a previous study of 
the effective mass from the Gell-Mann-Oakes-Renner relation.

\end{abstract}
\pacs{xxxxxx}

\narrowtext
Chiral symmetry and its explicit and/or spontaneous breakings 
are important aspects of strong interaction phenomenology. 
Chiral dynamics 
dominates the low-energy hadron structure and interactions. 
The chiral phase transition at finite temperature has been 
sought after experimentally for a long time. In addition, 
the weak interaction probes couple directly to the chiral 
currents, and the matrix elements of which sensitively 
depend on the chiral properties of hadron systems. On the 
theoretical frontier, however, massless fermions defy 
the naive nonperturbative treatments. Indeed, for more than
two decades, finding an appropriate fermion 
formulation has been one of the most difficult challenges 
in lattice quantum chromodynamics (QCD). In the 
last few years, Kaplan and Shamir's domain-wall 
construction \cite{kaplan,shamir} and Narayanan 
and Neuberger's overlap
fermion formalism \cite{overlap} have emerged as promising 
approaches to simulating massless quarks. In this paper, we 
aim to study the effectiveness of the domain-wall approach
which has already been used in a number of realistic numerical 
investigations \cite{blum,vranas,chen}. 

Following previous studies, we adopt 
Shamir's version of the domain wall fermion 
formulation \cite{shamir}, in which 
the five dimensional Wilson fermion is first introduced. 
The finite fifth dimension with $L_s$ lattice sites extends 
from $s=0$ to $s=L_s-1$. Dirichlet boundary condition 
on the quark fields is applied to the four-dimensional 
slices at $s=-1$ and $s=L_s$. The fifth component of the
gauge potential is identically zero and the other four
components are the same at every $s$ slice. The above 
construction in the gauge-field-free case yields
nearly-massless modes when the negative Wilson 
mass $m_0$ (the wall height) 
is taken between 0 and 10. These chiral fermions appear 
as the surface modes at $s=0$ and $s=L_s-1$ with opposite 
chirality. For $0<m_0<2$, one flavor Dirac
fermion can be constructed as 
\begin{equation}
     \psi(n) = P_+\psi(n, s=0) + P_-\psi(n, s=L_s-1) \ , 
\end{equation}
where $P_{\pm} = (1\pm \gamma_5)/2$ are the chiral 
projection operators and $n$ labels four-dimensional lattice
sites. For the finite wall separation ($L_s\ne\infty$), 
the chiral mode on the $s=0$ wall couples with 
the one with the opposite chirality
on the $s=L_s-1$ wall in an exponentially small 
way. Because of this coupling, a finite residual 
fermion mass is produced. The goal of 
this paper is to understand the size and dependence of this
induced fermion mass on $m_0$, $L_s$ when realistic 
background gauge fields are introduced.

\begin{figure}[t]
\begin{center}
\epsfig{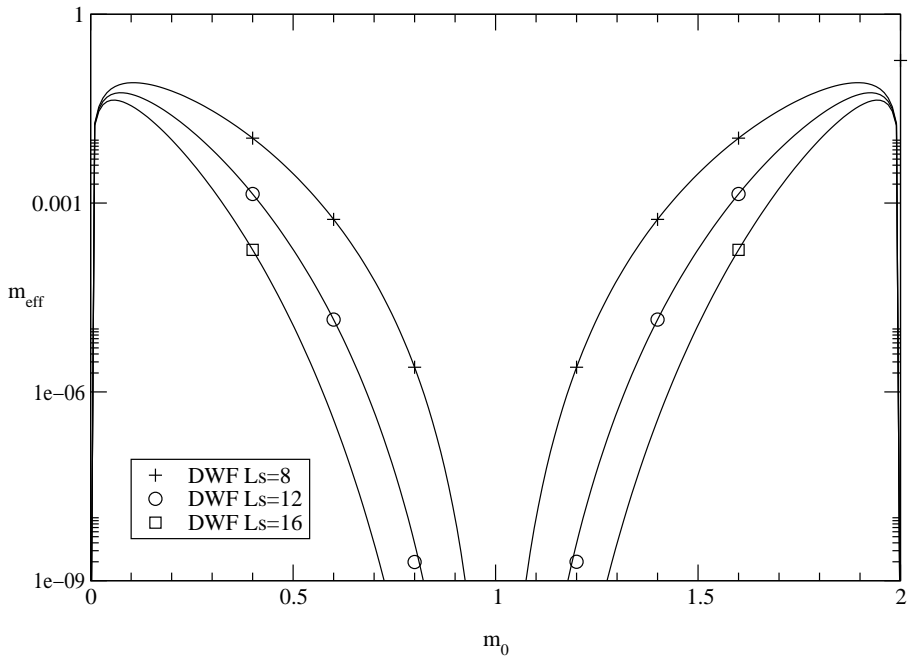}
\end{center}
\caption{Effective quark mass induced by domain-walls 
for the free field configuration. 
$L_s$ is the number of lattice sites in the fifth 
direction.}
\label{fig:1}
\end{figure}

In the absence of the gauge potentials, the effective mass can be
defined in terms of the pole of the free Green's function. 
For a large $L_s$, it has a simple analytical form \cite{shamir},
\begin{equation}
    m_{\rm eff} = m_0 (2-m_0) (1-m_0)^{L_s} \ . 
\end{equation}
One can also obtain $m_{\rm eff}$ by either  
diagonalizing the hermitian domain-wall Dirac operator 
$H_{DW}=\gamma_5 P_s D$ or $D^\dagger D = H^2_{DW}$ 
\cite{shamir,vranas}, 
where $D$ is the domain-wall Dirac operator and 
$P_s$ is the reflection along 
the fifth dimension. On a lattice with the periodic boundary 
condition, the lowest four-momentum of a fermion is zero, and 
the lowest eigenvalue of $\gamma_5 P_s D$ is just
$m_{\rm eff}$. To study the effective mass in an 
external gauge field, we have 
constructed a code to diagonalize $H_{DW}$ numerically. 
To test the code, we have computed $m_{\rm eff}$ 
on an $8^4$ lattice with $L_s=8,12,$ and 16, and  
the result is shown in Fig. \ref{fig:1}  together with the exact answer
from Eq. (1). An ideal massless fermion is obtained at 
$m_0=1$ where $m_{\rm eff}$ vanishes identically. 
For $m_0 \ne 1$, the induced quark mass decays exponentially
in $L_s$ with a rate depending strongly on $m_0$. 
The exponential decay slows down significantly 
as $m_0$ approaches 0 and 2. For 
any given accuracy $\epsilon$ and any $L_s$, there is a window 
surrounding $m_0=1$ in which $m_{\rm eff}<\epsilon$. 

In the presence of a realistic gauge potential, 
the effective quark mass result from the finite
wall separation may depend on how it is defined.  
Different definitions shall yield 
results consistent up to a factor of order 
unity. One approach is to exploit the explicit 
quark mass dependence in chiral Ward identities such as 
the Gell-Mann-Oakes-Renner (GMOR) relation as 
done in Ref. \cite{fleming}. 
Here we explore the effective mass in an alternative 
way. In continuum field theory, the Atiyah-Singer 
theorem\cite{at} states that the Dirac operator has 
a zero eigenvalue in the presence of an external 
background with topological charge $|Q|=1$. 
The explicit form of the solution 
was found by 't Hooft in 1976 \cite{thooft}. On the lattice, 
however, the notion of topological charge is ill defined:
any gauge configuration can be continuously deformed into
a null gauge field. Moreover, the discretization of 
an instanton field can introduce finite lattice-spacing effects 
lifting any exact zero eigenvalue. 
Therefore, a test of the Atiyah-Singer
theorem on lattice is usually complicated with
various lattice artifacts. 

There exists, however, a definition of lattice topology
and fermion zero mode which largely avoids this complication. 
In the overlap formalism, 
the Dirac operator is constructed from the overlap
of two many-fermion ground states \cite{overlap}. 
According to their recipe, one starts from a four-dimensional 
Wilson-Dirac operator with a negative Wilson mass $m_0$ and calculates 
its eigenvalues. For $m_0$ small 
and positive, the number of positive 
eigenvalues is equal to that of negative ones. When $m_0$ increases, a level
might cross from positive to negative or vice versa. 
When this happens, the gauge field is regarded 
to have a net topological charge $|Q|=1$. 
Then the overlap determinant
is exactly zero by construction. This definition
of lattice topology and zero mode do depend on, for
instance, the Wilson parameters $r$ and $m_0$. However,
the zero eigenvalue is exact, independent of the lattice
spacing $a$ and volume $V$. 

The domain-wall formulation can be regarded as
an approximation to the overlap formalism \cite{overlap}.  
Indeed, in the limit of $L_s\rightarrow \infty$, 
one recovers the overlap formalism apart from
some unimportant discretization effect 
in the fifth dimension. 
For a fixed gauge configuration and Wilson mass $m_0$,
if the overlap determinant is zero, $H_{DW}$
has an exact zero mode in the limit
of $L_s\rightarrow \infty$. 

In short, in a background gauge field, if the hermitian
Wilson-Dirac operator has a level crossing, 
the gauge field is considered to have a nontrivial topology. 
Edwards, Heller, and Narayanan have done extensive
studies of the topological properties of lattice gauge
configurations in this way \cite{edwards}. 
In a topologically-nontrivial background thus defined, 
the domain-wall Dirac operator with a finite $L_s$ has small
eigenvalues, nonvanishing only because of 
the finite wall separation. In the remainder of this
paper, we are mainly interested in such domain-wall 
effects on the fermion zero-mode. We {\it define} the smallest 
eigenvalues of the hermitian domain-wall operator 
$H_{DW}$ as the wall-induced effective fermion mass. 

\begin{figure}[hbt]
\begin{center}
\epsfig{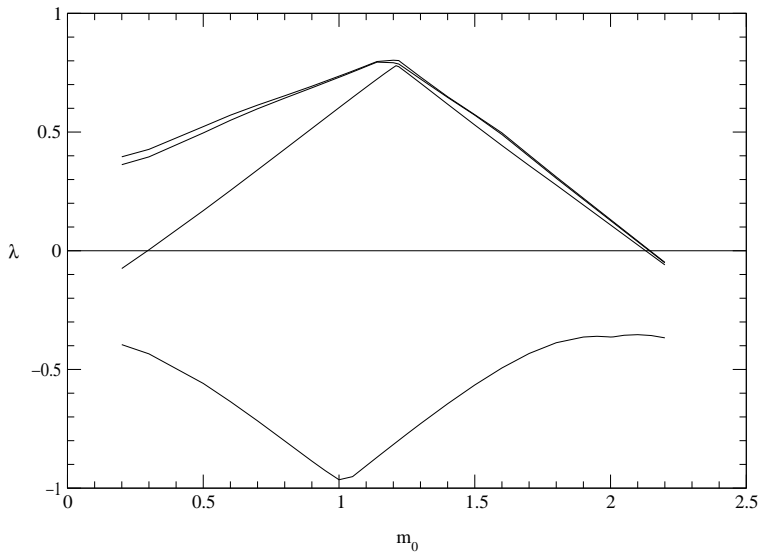}
\end{center}
\caption{Smallest eigenvalues of the hermitian 
Wilson-Dirac operator in a smooth 
(anti)instanton field described in the text.}
\label{fig:2a}
\end{figure}    

\begin{figure}[hbt]
\begin{center}
\epsfig{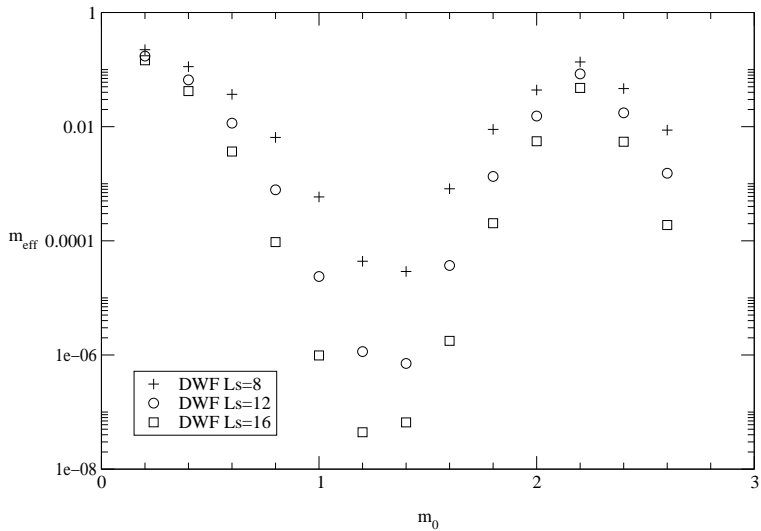}
\end{center}
\caption{Effective quark mass induced by domain walls 
in the same smooth instanton field as studied in Fig. 
\ref{fig:2a}. $L_s$ is the number of lattice sites in the fifth 
direction.}
\label{fig:2b}
\end{figure}    

As a first nontrivial example, we have shown in 
Figs. \ref{fig:2a} and \ref{fig:2b} the 
results in a smooth instanton field configuration on an $8^4$
lattice. A similar study has been reported in Refs. 
\cite{overlap,edwards,taiwantalk}. In our case,
the instanton configuration was generated according
to the prescription in Ref. \cite{chen2}: The size 
of the instanton $\rho_0$ is 10 and the cutoff 
parameter $r_{\rm max}$ is 3. The flow of 
the small eigenvalues of the hermitian Wilson-Dirac operator 
is shown in Fig. \ref{fig:2a}. A level crossing 
from positive to negative is seen at $m_0 = 0.3$. 
Four separate crossings in the opposite 
direction happen near $m_0=2.2$. Only three of the crossings are plotted. 
In the overlap formalism, the Dirac operator 
has an exact zero eigenvalue in the region between the crossings. 
 
The lowest eigenvalue of the hermitian domain-wall 
Dirac operator $H_{\rm DW} = \gamma_5 P_s D$ is 
shown in logarithmic scale in Fig. \ref{fig:2b}. 
The overall profile of the eigenvalue 
as a function of $m_0$ is similar to 
the free case in Fig. \ref{fig:1}. For a fixed $L_s$, 
the smallest eigenvalue occurs at 
around $m_0 = 1.3$, shifted upward from 
$m_0 = 1$. This shift reflects the renormalization 
of the Wilson mass in the presence of the external 
gluon field. As $m_0$ deviates from $m_0= 1.3$, the 
domain-wall effects grow stronger.
For a fixed $m_0$, the effective fermion
mass decreases exponentially as $L_s$ 
increases from 8 to 12 and 12 to 16.

\begin{figure} [hbt]
\begin{center}
\epsfig{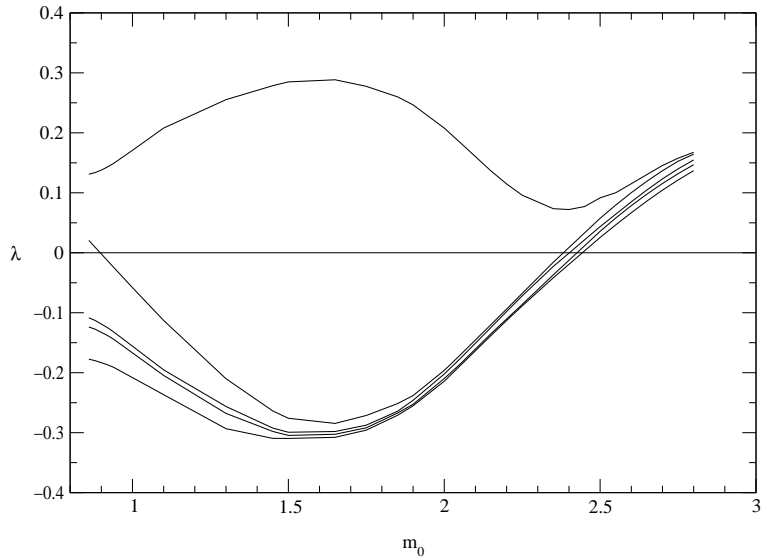}
\end{center}
\caption{Eigenvalues of the hermitian 
Wilson-Dirac operator as a function of the 
negative Wilson mass $m_0$ in a topological 
gauge configuration generated on an $8^4$ lattice 
and at $\beta=6.0$.}
\label{fig:3a}
\end{figure}    

\begin{figure} [hbt]
\begin{center}
\epsfig{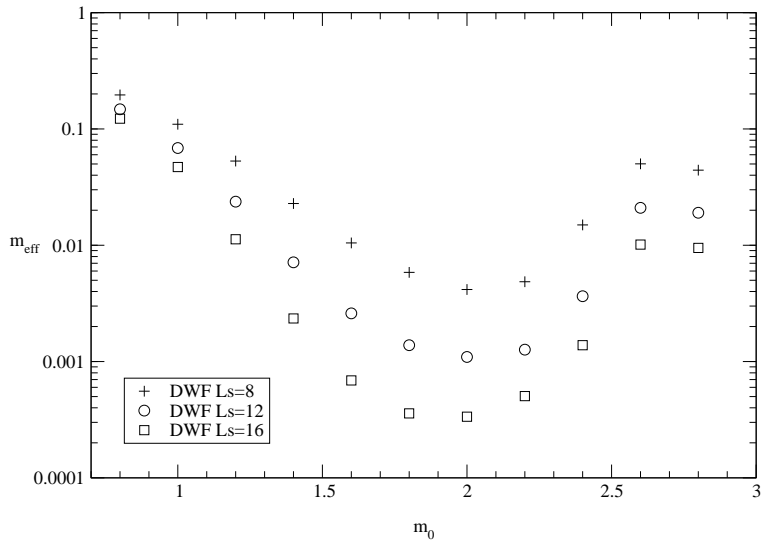}
\end{center}
\caption{Effective quark mass induced by the domain walls
in the same topological configuration as 
studied in Fig. \ref{fig:3a}.}
\label{fig:3b}
\end{figure}    

\begin{figure} [hbt]
\begin{center}
\epsfig{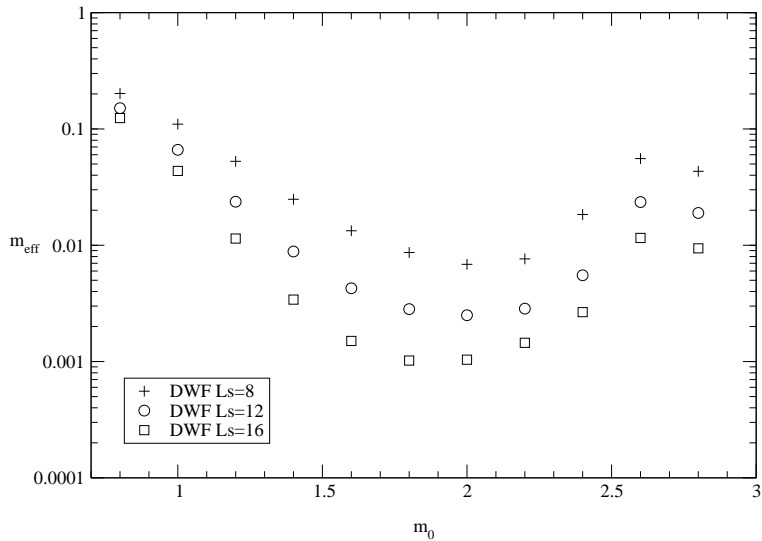}
\end{center}
\caption{Same as Fig. \ref{fig:3b}, but 
for a different topologically nontrivial configuration.}
\label{fig:3c}
\end{figure}    

Our result is quantitatively consistent with the chiral 
condensate $\langle \overline{\psi}\psi\rangle$
calculation in Ref. \cite{chen}. For 
instance, at $L_s=10$ and $m_0=1.2$, we have
$m_{\rm eff} \sim 7\times 10^{-6}$. We expect then 
$\langle \overline{\psi}\psi\rangle$ grows like $1/m_f$ 
as the explicit quark mass parameter $m_f$ reduces 
to $\sim 10^{-5}$. 
For a fixed $m_f=5\times 10^{-4}$ 
and $L_s=10$, there is a window in $m_0\sim[0.9,1.6]$
in Fig. \ref{fig:2b} where the induced quark 
mass $m_{\rm eff}$ is smaller than $m_f$.  
Hence we expect $\langle \overline{\psi}\psi\rangle$
is approximately independent of $m_0$ there. 
At $m_0=0.75$, the effective quark mass 
is about $10^{-3}$ with $L_s=12$. We expect the 
chiral condensate to have little sensitivity on $m_f$ 
when $m_f>10^{-3}$. All of the above are  
in accordance  with those reported
in Ref. \cite{chen}. 

A smooth instanton on the lattice is far from a 
typical equilibrium gauge configuration entering in 
the Feynman path integral. 
A more realistic study of the 
induced quark masses requires 
gauge configurations with quantum 
fluctuations fully included. In the 
following, we work on a set of Monte 
Carlo configurations generated on
a four-dimensional $8^4$ lattice and with  
$\beta=6$. The physical volume is 
somewhat small, but we suspect that the
domain-wall effects have a weak dependence
on it.

We pick a lattice configuration in which the lowest
eigenvalue of the hermitian Wilson-Dirac operator
crosses from the positive to the negative 
at $m_0$ near 0.9. In Fig.\ref{fig:3a}, we have shown 
the flow of the lowest few eigenvalues as a function of $m_0$. 
Between $m_0=2.35$ and $2.45$, four levels cross  
from the negative to positive region. 
Although not shown explicitly, six level crossings
occur at around $m_0=4$. As pointed out in Ref. \cite{edwards}, 
the hermitian Wilson-Dirac operator is symmetric with respect
to $m_0=4$ and hence the flow diagram has the same symmetry.
According to the overlap fermion formalism, 
the Neuberger-Dirac operator in the above 
gauge background has one zero 
eigenvalue when $m_0$ is in the interval $[0.9, 2.35]$, 
three zero eigenvalues in $[2.45, 4]$, three 
again in $[4, 5.55]$, and finally one 
in $[5.65,7.1]$. 

In Fig.\ref{fig:3b}, we have shown the smallest eigenvalue
of the hermitian domain-wall Dirac operator $H_{DW}$. 
Different symbols correspond to three 
different $L_s=8$ (pluses), $12$ (circles), 
and $16$ (squares). Over a large region 
of $m_0$, the effective quark mass decreases 
exponentially in $L_s$, as is clear 
from the approximate equal spacings between
pluses, circles, and squares. The fastest 
decay occurs at $m_0$ around $2.0$, compared
with 1.0 in the free case and 1.3 in the smooth 
instanton field. The wall effects become 
strong again near $m_0=2.7$ 
beyond which we find four small eigenvalues (not shown). 
To our surprise, this transition point to  
the doubler region is at higher $m_0$  
compared with the prediction from the 
spectral flow of the hermitian Wilson-Dirac 
operator. This may indicate some subtle differences
between the eigenvalues of the transfer 
matrix in the domain-wall formalism and those 
of the hermitian Wilson-Dirac operator 
in the overlap formalism.

The important point about Fig.\ref{fig:3b} is that the 
magnitude of the effective mass is much enhanced relative to 
the case of the smooth instanton configuration. 
With a best choice of $m_0$ near 2.0, 
the domain-wall Dirac operator has the smallest
eigenvalue $(3\sim 4)\times 10^{-4}$ at $L_s=16$. 
For the same $L_s$ and with $m_0$ of 1.3, 
the smooth instanton configuration yields an eigenvalue
$\sim 10^{-8}$. This dramatic increase of the effective mass
 comes from the ultraviolet 
fluctuations. As we will discuss further below,
the ultraviolet fluctuations at 
strong coupling can cause great trouble
for the domain-wall formalism. In Fig. \ref{fig:3c}, 
we have shown the smallest eigenvalue of $H_{DW}$
in another topological
gauge configuration. The result is 
qualitatively similar to that in Fig. \ref{fig:3b}. 
In particular, for $L_s=16$ and $m_0=2.0$, the induced
quark mass is around $10^{-4}$. 

Since the effective quark mass 
depends on particular gauge fields, it is useful to
get an average over an ensemble of gauge 
configurations. For this purpose, we have generated a 
set of 150 configurations on a $8^4$ lattice 
at $\beta =6.0$. By studying the spectral 
flow of the hermitian Wilson-Dirac 
operators, we found 12 topologically-nontrivial 
configurations. The second column in Table I
shows $m_0$ at which the first level crossing occurs.
All the crossing points are entirely concentrated
in the interval between 0.8 and 1, a fact 
consistent with a similar study in a 
slightly larger lattice $8^3\times 16$ 
\cite{edwards}. We proceed to calculate the 
lowest eigenvalue of the hermitian domain-wall
Dirac operator with $L_s=8,12,$ and $16$ at 
$m_0=1.8$. The result is shown as the remaining 
columns in Table 1. We find
the average effective quark mass at $L_s=16$ is
$8(4) \times 10^{-4}$. The dependence of 
the average effective mass in $L_s$ is consistent
with the exponential within the error.

Our result can be compared with a previous study of the
wall-induced quark mass using the Gell-Mann-Oakes-Renner
relation on a $16^3\times 32$ lattice at the
same value of $\beta$ \cite{fleming}. In the domain-wall
formulation, the quenched chiral condensate
is related to the pion susceptibility $\chi_\pi$ 
\cite{fleming} by
\begin{equation}
  \langle \overline{\psi}\psi\rangle
  = (m_f+m_{\rm eff}^{\rm GMOR}) \chi_\pi + b
\end{equation}
where $b$ is a constant vanishing in the 
$L_s\rightarrow \infty$ limit. 
The effective quark mass $m_{\rm eff}$ can be 
extracted from $m_f$-dependence of the condensate 
and $\chi_\pi$. For $L_s=16$ and $m_0=1.8$, 
an extrapolation to $m_f=0$ limit yields 
$m_{\rm eff}^{\rm GMOR} = 0.012$. Note that 
the smallest $m_f$ at which the data is taken  
is $0.01$.

To understand the physical significance
of $m_{\rm eff}^{\rm GMOR}$, we recall the spectral
version of the GMOR relation in the overlap
formulation \cite{edwards1}:
\begin{equation}
  \langle \overline{\psi}\psi\rangle
  = m_f\chi_\pi = 
m_f\left({|Q|\over m_f^2 V}
   + {1\over V} \sum_i {2(1-\lambda_i^2)
  \over \lambda_i^2 + m_f^2} \right)\ , 
\end{equation}
where $\pm\lambda_i$ are a conjugating pair of eigenvalues of 
$(\gamma_5 D_{\rm ov})^2$($D_{\rm ov}$ = overlap Dirac operator). 
The first term signifies the contribution
from the topological zero modes; for sufficiently 
large volume V and/or large $m_f$, 
this topological charge term is negligible. 
The second term measures the chiral symmetry 
breaking effects on the conjugating 
pairs of eigenvalues. In the study quoted above 
\cite{fleming}, the topological charge term 
is insignificant even at the smallest $m_f=0.01$,   
and the extracted $m^{\rm GMOR}_{\rm
eff}$ undoubtly measures the explicit chiral-symmetry
breaking effects in conjugating pairs of eigenvalues
as induced by the domain walls.

One can push the quenched calculation in Ref. \cite{fleming}
to the limit $m_f=0$. In this case, both the chiral 
condensate $\langle \overline{\psi}\psi\rangle$ and 
pion susceptibility are dominated by the zero
modes. The quark mass obtained from the ratio 
of the two observables is just the smallest eigenvalue
of the Dirac operator in the topologically-nontrivial
configurations. Therefore in the quenched chiral limit,
the effective quark mass determined from 
the GMOR relation coincides with what we have considered 
in this paper. To be sure, the effective 
mass defined from the effects on the fermion zero modes 
is not the same as the one defined from the effects on
the conjugating pairs of eigenvalues. Nonetheless, 
both definitions shall be consistent within
a factor of order unity. In this spirit, our average
effective quark mass $8(4)\times 10^{-4}$ is indeed in 
accordance with $1.2\times 10^{-3}$ from Ref. \cite{fleming}.

To be completely sure about the consistency of the 
two approaches, further numerical studies are needed.
For instance, along the line of study in Ref. \cite{fleming}
one can attempt to subtract the zero-mode contribution
to the chiral condensate and pion susceptibility, and
then both quantities can be measured in the 
$m_f\rightarrow 0$ limit. On the other hand, the present 
study can be repeated at a larger physical volume; 
a $16^3\times 32$ lattice will be more suitable 
for comparison.

\begin{table}[hbt]
\begin{center}
\begin{tabular}{ccccc}
Configuration Number & Crossing in $m_0$& $m_{\rm eff}(L_s =8)$ &$m_{\rm eff}(L_s=12)$&$m_{\rm eff}(L_s=16)$\\
\hline
14    &   0.916	&	6.777e-03	&	1.745e-03	&	5.014e-04\\
38    &   0.874	&	8.663e-03	&	2.822e-03	&	1.019e-03\\
42    &   0.876	&	6.104e-03	&	1.694e-03	&	5.601e-04\\
45    &   0.897	&	5.849e-03	&     1.382e-03	&     3.584e-04\\
58    &   0.945	&	9.188e-03	&	2.607e-03	&	8.037e-04\\
82    &   0.862	&	6.356e-03	&	1.718e-03	&	5.298e-04\\
92    &   0.901	&	1.002e-02	&	3.619e-03	&	1.471e-03\\
98    &   0.853	&	5.543e-03	&	1.432e-03	&	4.189e-04\\
101   &   0.842	&	5.948e-03	&	2.395e-03	&	1.449e-03\\
129   &   0.845	&	4.525e-03	&	1.094e-03	&	3.066e-04\\
133   &   0.877	&	7.554e-03	&	2.276e-03	&	7.677e-04\\
147   &   1.003	&	1.252e-02	&	4.091e-03	&	1.447e-03\\
\hline
$\overline{m_{\rm eff}}$&& 0.0074(22)&0.0022(9)&0.0008(4)\\
\end{tabular}
\end{center}
\caption{ Average effective mass for the topologically nontrivial configurations}
\end{table}

We finally return to the central question of the
domain-wall fermion formalism: How large an $L_s$
is needed for a practical simulation? Clearly, 
the results for the free field  or artificially 
smooth gauge fields are of no help here. The answer
depends on the size of quantum fluctuations. For large
values of $\beta$ (and perhaps small physical volumes as well),
such as the case we have presented, 
one can work with $L_s = 12$ or 16 and 
keep the induced quark mass under control. 
However, in going to smaller $\beta$ and larger 
physical volume, level crossing happens 
continuously in a region of $m_0$ 
above some critical value \cite{edwards}. From Figs. \ref{fig:3b}
and \ref{fig:3c}, we expect that if a level 
crossing occurs slightly before the $m_0$ where the 
domain-wall Dirac operator is defined, 
the wall-induced quark mass will be huge. [When the level 
crossing happens slightly above $m_0$, 
the chiral symmetry breaking effects in the 
conjugation pairs of eigenvalues is expected to be 
strong. Again this leads to a large
effective quark mass.] Then the average 
effective quark mass can be strongly influenced by  
this type of accidental configurations depending on 
the frequency they occur. The level crossings 
at large $m_0$ reflect strong quantum fluctuations
at the scale of the lattice spacing \cite{edwards}. 
Therefore, it is not surprising that at $\beta=5.7$, one needs
to have very large $L_s$ (30 to 40) to 
keep $m_{\rm eff}$ small; of course, this is the price
that one has to pay to keep the physical volume
large.

To summarize, we have studied the induced quark mass
resulted from the finite domain wall separation by 
diagonalizing the hermitian domain-wall Dirac operator 
in topologically nontrivial configurations. We find the 
quantum fluctuation strongly enhances the domain-wall effects. 
However, the effective mass does show an exponential decay 
as a function of $L_s$. Our result on an $8^4$ lattice 
with $\beta=6$ is consistent with the effective fermion masses
from the GMOR relation, although a detailed analysis 
shows that the two definitions of the effective mass
are not the same. Finally, we comment on the size
of $L_s$ needed in a practical Monte Carlo simulation.

\acknowledgements
We thank N. Christ, R. Edwards, and J. Negele for useful
discussions related to the subject of this paper. The 
numerical calculation reported here
was performed on the Calico Alpha Linux Cluster at 
the Jefferson Laboratory, Virginia.
This work is supported in part by funds provided by the
U.S.  Department of Energy (D.O.E.) under cooperative agreement
DOE-FG02-93ER-40762.

\appendix
\section{EIGENVALUES OF FREE DOMAIN-WALL FERMION}
\narrowtext
The domain-wall induced fermion mass in the free
case was first calculated by Shamir \cite{shamir} using
Green's function approach. Vranas stated in his
paper \cite{vranas} that he obtained the same result 
by diagonalizing the domain-wall Dirac operator
without showing the actual calculation. 
An explicit derivation of $m_{\rm eff}$ in the $m_f=0$
case was later provided by Neuberger \cite{neubergermass}. 
Here we show a complete derivation with 
the inclusion of $m_f$.

The domain-wall Dirac operator in the free field limit in 
momentum space can be written as:

\begin{eqnarray}
&D = i\bar p +M^+ P_++ M^- P_{-}\ ,
\end{eqnarray}
where  $P_{\pm} = \frac12(1 \pm \gamma_5)$ and ${\bar p}= \gamma_{\mu} \sin p^{\mu} $.
Mass matrices $M^{\pm}$ are defined as
\begin{eqnarray}
&(M^+)_{ss'} = \delta_{s+1,s'} -b(p)\delta_{s,s'} -m_f\delta_{s,N_s}\delta_{s',1}& \ ,\nonumber \\
&(M^-)_{ss'} = \delta_{s-1,s'} -b(p)\delta_{s,s'} -m_f\delta_{s,1}\delta_{s',N_s}&\quad\ (s, s'=1,\cdots N_s)\ ,
\end{eqnarray}
where $b(p) = 1-m_0 +\sum_{\mu} (1-\cos p_{\mu})$ and $m_f$ is the explicit fermion mass. 
Our goal is to calculate the smallest eigenvalue of the bilinear hermitian domain-wall Dirac
operator
\begin{eqnarray}
&DD^{\dagger} = {\bar p}^2 +M^+ M^{-} P_+  + M^- M^+ P_-.
\end{eqnarray}
Since $M^+M^-$ and  $M^-M^+$ have the same eigenvalue spectrum,
it is sufficient to consider
\begin{eqnarray}
M^+ M^- = \left(
\begin{array}{lllll} 
b^{2}+1 & - b &0 &\cdots &m_fb\\
 -b   & b^2+1& -b&\cdots & 0\\
\vdots & & &\ddots & \\
m_fb& \cdots&0 & -b&b^2+m_f^2
\end{array}\right).
\end{eqnarray}
The second to $(N_s-1)\ $-th row of the secular equation
$(M^+ M^-)_{ss'} \Psi_{s'} = \lambda^2 \Psi_s$, or,
\begin{eqnarray}
(b^2+1)\Psi_{s} -b(\Psi_{s-1}+\Psi_{s+1})=\lambda^2\Psi_s,
\end{eqnarray}
 is solved by
$\Psi_s =  \exp [\pm \alpha s]\ (s= 1, \cdots N_s)\ ,$ provided $\lambda$ and $\alpha$ satisfy
\begin{eqnarray}
-2b \cosh(\alpha) + (b^2+1-\lambda^2) = 0. \label{eq:lambda}
\end{eqnarray}
The first and the last rows of the secular equation can be satisfied by a linear combination of exponential solutions,
$\Psi_s = \exp[-\alpha(s-1)] +A \exp[-\alpha(N_s-s)]\ $, where $A$ is a constant to be determined:
\begin{eqnarray}
&(b^2+1 -\lambda^2)(1+Ae^{-\alpha(N_s-1)}) -b(e^{-\alpha}+A e^{-\alpha(N_s-2)}) +m_fb (e^{-\alpha(N_s-1)} +A)=0\ , \nonumber \\
&(b^2+m_f^2 -\lambda^2)(e^{-\alpha(N_s-1)}+A) -b(e^{-\alpha(N_s-2)}+A e^{-\alpha}) +m_fb (1+ A e^{-\alpha(N_s-1)} )=0\ . 
\end{eqnarray}
Using Eq. (\ref{eq:lambda}) to eliminate $\lambda$ from the above, we get
\begin{eqnarray}
2\cosh(\alpha)(1+Ae^{-\alpha(N_s-1)}) -(e^{-\alpha}+A e^{-\alpha(N_s-2)}&&) +m_f (e^{-\alpha(N_s-1)} +A)=0, \nonumber \\
(2\cosh(\alpha)-(1-m_f^2)/2b)(e^{-\alpha(N_s-1)}+A)&& \label{eq:A}\\
-(e^{-\alpha(N_s-2)}&&+A e^{-\alpha}) +m_f (1+ A e^{-\alpha(N_s-1)} )=0. \nonumber
\end{eqnarray}
Eliminating A from Eq. (\ref{eq:A}) and rearranging terms, we have
\begin{eqnarray}
e^{2\alpha} -e^{\alpha}/b -m_f^2(1-e^{\alpha}/b)& +2m_fe^{-\alpha N_s}( e^{2\alpha} -1)& \nonumber \\
&+m_f^2e^{-\alpha(2N_s-2)}(e^{\alpha}&-1/b)-e^{-\alpha 2N_s}(1-e^{\alpha}/b)=0.
\end{eqnarray}
In the $N_s= \infty, m_f=0$ limit, $e^{\alpha}=1/b$.
Assuming  $e^{\alpha}=1/b + \delta$ and keeping terms linear in
$\delta$, we find
\begin{eqnarray}
&\delta \sim -(m_f+b^{N_s})^2(1-b^2)/b.
\end{eqnarray}
Finally, substituting $e^{\alpha}=1/b + \delta$ into Eq. (\ref{eq:lambda}), we get the eigenvalue,
\begin{eqnarray}
&\lambda^2 = b^2+1-b(e^{\alpha}+e^{-\alpha}) \sim -b(1-b^2)\delta = (1-b^2)^2 (m_f+b^{N_s})^2\ , \nonumber
\end{eqnarray}
or $\lambda = (1-b^2)(m_f+b^{N_s})$, as quoted in Ref. \cite{shamir}.

\end{document}